\documentclass[showpacs,twocolumn,preprintnumbers,amsmath,amssymb,nofootinbib]{revtex4-1}
\usepackage{stmaryrd}
\usepackage{color}
\usepackage{mathrsfs}
\usepackage{rotating}
\usepackage{graphicx} 
\usepackage{amsfonts}
\usepackage{mathrsfs}
\usepackage[american]{babel}
\usepackage{pstricks}
\usepackage{pst-plot}
\usepackage{pst-grad}
\usepackage{pst-eps}
\usepackage{slashed}

\newcommand{\bea}{\begin{eqnarray}}
\newcommand{\eea}{\end{eqnarray}}
\newcommand{\beq}{\begin{equation}}
\newcommand{\eeq}{\end{equation}}

\begin{document}
\title{Dual fermion condensates in curved space}
\author{Antonino Flachi}
\affiliation{Centro Multidisciplinar de Astrof\'{\i}sica,
Departamento de F\'{\i}sica, Instituto Superior T\'ecnico,
Universidade T\'{e}cnica de Lisboa,\\ Avenida Rovisco Pais 1, 1049-001
Lisboa, Portugal}
\date{\textcolor{blue}{\today}}

\begin{abstract}
In this paper we compute the effective action at finite temperature and density for the dual fermion condensate in curved space with the fermions described by an effective field theory with four-point interactions.
The approach we adopt refines a technique developed earlier to study chiral symmetry breaking in curved space and it is generalized here to include the U$(1)$-valued boundary conditions necessary to define the dual condensate. The method we present is general, includes the coupling between the fermion condensate and the Polyakov loop, and applies to any ultrastatic background spacetime with a nonsingular base. It also allows one to include inhomogeneous and anisotropic phases and therefore it is suitable to study situations where the geometry is not homogeneous. We first illustrate a procedure, based on heat kernels, useful to deal with situations where the dual and chiral condensates (as well as any smooth background field eventually present) are slowly or rapidly varying functions in space. Then we discuss a different approach based on the density of states method and on the use of Tauberian theorems to handle the case of arbitrary chemical potentials. As a trial application, we consider the case of constant curvature spacetimes and show how to compute numerically the dual fermion condensate in the case of both homogeneous and inhomogeneous phases.
\end{abstract}
\maketitle

\section{Introduction}

Understanding the relation between confinement and chiral symmetry breaking in quantum chromodynamics (QCD) remains as a challenging problem in theoretical particle physics. The difficulty to relate the two phenomena is partly due to their nonperturbative nature, partly because they occur in different sectors of the theory. While chiral symmetry resides in the quark sector, with the chiral condensate acting as an order parameter for chiral phase transitions with vanishing quark masses, confinement is a property linked with the gluon sector, with the Polyakov loop being an order parameter, for infinitely large quark masses, signaling the breaking of the center symmetry at the deconfinement transition.
For chiral symmetry the situation is relatively clear, owing to the Banks-Casher formula \cite{banks} that connects the chiral condensate with the eigenvalue density of the Dirac operator suggesting a nonperturbative mechanism inducing chiral symmetry breaking below a critical temperature \cite{shuryak,diakonov}. For confinement an analogously simple explanation is presently not known.

Lattice QCD simulations at finite temperature are not yet conclusive \cite{mistery}. If, on the one hand, they suggest the existence of a specific mechanism relating the two phenomena, on the other hand, predictions for the critical temperatures at which the two transitions occur differ. While some lattice results indicate a sizable difference in the critical temperatures at which chiral symmetry breaking and deconfinement occur \cite{budapest}, others \cite{bielefeld} point to the same value for both transitions. 

A step forward in trying to understand how chiral symmetry and confinement may be related has been taken by Gattringer and collaborators who started to explore possible ways to connect quantities sensitive to confinement with the spectral properties of the Dirac operator \cite{gattringer,erek}. By looking at how spectral sums for the Dirac operator transform under center symmetry, they argued that quark confinement is encoded in the low-lying spectral modes, suggesting a direct way to connect chiral symmetry with confinement. A possible way to construct an order parameter for the center symmetry was given explicitly in Ref.~\cite{erek} by taking the Fourier transform of the quark condensate with respect to a set of phase dependent boundary conditions for the fermions. By means of this transformation the quark condensate turns into the expectation value of an equivalence class of Polyakov loops with winding number $n\in \mathbb{Z}$ conjugate to the phase. This quantity has been termed dual condensate or dressed Polyakov loop. Details will be given in the next section. 

This idea has attracted considerable attention for various reasons. First of all, it is physically transparent, since spectral quantities provide a natural decomposition into IR and UV parts, therefore allowing for the possibility to study how individual parts of the spectrum affect the transition to deconfinement. Secondly, it provides a precise and simple computational scheme relating chiral and center symmetry: not only the dual condensate can be straightforwardly computed numerically on the lattice \cite{erek}, but can also be studied by means of analytic and functional methods \cite{fischer}. A third important reason, relevant to the present discussion, is that it naturally fits effective descriptions of QCD allowing one to consider situations where a lattice approach may be more challenging.
In fact, the dual quark condensate has been originally introduced in the context of lattice QCD, with the actual computation performed in the quenched approximation \cite{erek}. Within this scheme, the U$(1)$-valued condensate is averaged over the gauge field configurations with the boundary conditions imposed on the fermions {\it only}, while gauge degrees of freedom respect the usual periodic boundary conditions.

On the other hand, within effective descriptions of QCD, {\it i.e.}, {\it \`a la} Nambu-Jona Lasinio \cite{njl} (see Refs.~\cite{review1,review2,review3} for reviews), the gluon mediated interactions are encoded in the effective interactions among quarks and in order to have a way to model (an order parameter for) confinement, it is necessary to introduce also the effects of the gauge fields in some explicit way. A known extension of this sort is the Polyakov extended Nambu-Jona Lasinio model (see \cite{fukushima}). In fact, within such construction, an order parameter to model confinement, {\it i.e.}, the Polyakov loop, would already be available. However, extending the model of \cite{fukushima} to curved space would encounter at least one serious difficulty in relation to the fact that the purely gluonic part of the potential is fitted from lattice data that are, to the best of our knowledge, not available on curved backgrounds. This limitation can be partially overcome by using the dual quark condensate as order parameter as long as the coupling of the chiral condensate with the Polyakov loop is appropriately introduced. 
Although the Polyakov loop can only be incorporated through some phenomenological parametrization (fixing completely the Polyakov loop would require some direct, numerical computation of the purely gluonic potential potential in curved space at strong coupling), still the dual condensate can be calculated as a general function of the Polyakov loop. 

The goal of this paper is to illustrate how to compute the dual fermion condensate in the context of a four-fermion effective theory with the Polyakov loop coupled to the fermion condensate on a curved $D$-dimensional ultrastatic Riemannian space that we assume to be nonsingular and without boundaries. While analyses of the dual fermion condensate within the Nambu-Jona Lasinio model in flat spacetime have appeared in the literature (see, for example, Refs.~\cite{Kashiwa:2009ki,ruggieri,mukherjee,Xu:2011pz}), analogous studies in curved space are very limited. The only work we are aware of is that of Ref.~\cite{sasagawa}, where the interesting question of how geometrical effects influence confinement is addressed in a precise and computationally accessible framework. Specifically, Ref.~\cite{sasagawa} presents a computation of the dual fermion condensate for background geometries of the form $\mathbb{R} \otimes \mathbb{S}^3$ and $\mathbb{R} \otimes \mathbb{H}^3$ where the base spatial sections $\mathbb{S}^3$ and $\mathbb{H}^3$ define, respectively, a three-dimensional sphere and a three-dimensional hyperbolic space. 

Our approach (and related computation) will differ from the work of \cite{sasagawa} in several ways. First of all, the class of geometries considered in Ref.~\cite{sasagawa} was restricted to a specific choice of the base as described above. Here, we only make the assumption of ultrastaticity and smoothness of the geometry, but otherwise keep ourselves general. Secondly, the method adopted in \cite{sasagawa} relies on a direct computation of the partition function that requires explicit knowledge of the eigenvalues (or eigenvalue density) of the Dirac operator. This is possible for the two cases considered in Ref.~\cite{sasagawa}, but not in general. On the other hand, the approach we will follow here allows one to overcome this limitation and can be used also in situations where the spectrum is not explicitly known. Although we assume the background geometry to be ultrastatic and boundaryless, generalizations that include static spacetimes and boundaries can be straightforwardly performed starting from the analysis we present here. (General considerations and related work regarding chiral symmetry breaking in curved space can be consulted, for instance, in Refs.~\cite{sergei,forkel,klim}. Effects of boundaries on chiral symmetry breaking  have been considered, for example, in Refs.~\cite{ek,fl1,fl2,tib}.) Practically, we elaborate on our previous work (see Ref.~\cite{flachitanaka}), where the effective action for the chiral condensate was calculated using heat-kernel and zeta-regularization techniques, and extend it here to include U$(1)$-valued boundary conditions. We shall show that, while the general structure and factorization property of the zeta function will not change, the dependence on the boundary conditions modifies the thermodynamical kernel introducing an explicit dependence on the phase. 

Another important point where our approach departs from the direct method of Ref.~\cite{sasagawa} is that it allows for a direct inclusion of inhomogeneous phases. In flat space, for small chemical potentials, this difference only becomes relevant to characterize excited states, since the ground state is expected to be homogeneous. The same is true when the background spacetime has constant curvature and no boundaries. However, for more general situations, as, for instance, in the case of black holes, in the presence of boundaries or for topologically nontrivial geometries, both the dual and the chiral fermion condensates {\it will not} be homogeneous. For these more general cases, a modification like the one used here becomes necessary also to characterize the features of the ground state. 

Finally, unlike Ref.~\cite{sasagawa}, we include in our treatment an explicit coupling between the Polyakov loop and the chiral condensate. 

Details of the model and the basics of the method will be described in the next section where we explicitly construct the effective action for the dual fermion condensate on a generic smooth curved background with compact spatial section. 
The scheme adopted, analogously to what we have done for chiral condensates, uses a quasi-nonperturbative ansatz for the heat trace. This allows one to resum all powers of the scalar curvature and of the dual fermion condensate and implicitly assumes that spatial variations do not occur `rapidly' over space. If one wishes to consider cases in which variations do occur rapidly, although the basics of the method will not change, a different ansatz for the heat trace, like the one described in \cite{gusev}, should be used after appropriate generalization to include a nonvanishing chemical potential. This will be briefly discussed in the Appendix \ref{app}, but not implemented numerically. In Sec.~\ref{sec3} we will describe a different approach to compute the effective action for the dual condensate based on the density of states method and on the use of Tauberian theorems. 

Although the approach of this paper provides a tool to study a setup similar to that of Ref.~\cite{sasagawa}, our main goal is to prepare the formal machinery to study more complex situation of black holes and to test it in a relatively simple case. Such an example will be described in Sec.~\ref{sec4}, where we will consider the case of a constant curvature geometry. Concretely, we will apply the formal results described in the previous sections and, by means of numerical analysis, we will construct the dual fermion condensate for both homogeneous and inhomogeneous phases. Our conclusions will close the paper.

\section{Dual Fermion Condensate in Curved Space}
\label{sec2}

We will begin this section describing the setup. The basics are the same as in Ref.~\cite{flachitanaka} that the reader is invited to consult for additional details. Notation follows Ref.~\cite{flachitanaka} and Planck units are used everywhere. 

The action for the model under consideration takes the form
\bea
S = \int d^Dx \sqrt{g} \left\{ 
\bar \psi i \gamma^\mu \nabla_\mu \psi 
-m\bar\psi\psi
+ 
{\lambda\over 2N}
\left(\bar \psi \psi\right)^2 
\right\},
\label{eqt2.1}
\eea
with $\psi$ being a Dirac spinor, $N$ 
the number of fermion degrees of freedom, $g$ the determinant of the metric and $\lambda$ the coupling constant. The background geometry is assumed to be $D=d+1$ dimensional and ultrastatic with line element of the form 
\bea
ds^2 = dt^2 - g_{ij} dx^i dx^j ~,
\label{geometry}
\eea
with $g_{ij}$ being the metric on the $d$-dimensional spatial section of the spacetime. The covariantization of the model has been discussed in Ref.~\cite{flachitanaka} (see also the review \cite{sergei}) and we will not repeat it here. In the present treatment we will keep things as simple as possible, and, thus, neglect all nonminimal couplings between matter and gravity as well as terms of the form $\left(\bar \psi\imath \gamma^5 {\bf \tau} \psi\right)^2$ and higher order ones. We remark, however, that dealing with these terms does not require any formal changes with respect to those presented here. The Nambu-Jona Lasinio model reduces to (\ref{eqt2.1}) when the pseudoscalar term is suppressed.

The goal of this section is to compute the effective action for the dual quark condensate $\Sigma_n$ with winding number $n\in \mathbb{Z}$. This is defined in terms of the fermion condensate, as the scalar expectation value of the fermion bilinear with U$(1)$-valued boundary conditions
\bea
\Sigma_n =\int_0^{2\pi} {d\varphi\over 2\pi} {e^{- i \varphi n}} \langle \bar\psi \psi\rangle_{\varphi}~,
\label{dual}
\eea
where the expectation value $\langle \bar\psi \psi\rangle_{\varphi}$ is calculated imposing the following generalized boundary conditions along the temporal direction
\bea
\psi (x_i, \beta) = e^{-i \varphi} \psi (x_i, 0)~,
\label{bcs}
\eea
with $\varphi \in \left[ 0, 2\pi \right)$ and $\beta=1/T$. For $\varphi = \pi$, $\langle \bar\psi \psi\rangle_{\varphi=\pi}$ returns the standard fermion condensate.  
The above quantity (\ref{dual}) has been introduced in Ref.~\cite{erek} where it was shown, in the context of lattice QCD, that can be expanded as a series of closed loops, with the winding number $n$ representing the number of times the loop winds around the compact time direction. For $n=1$, $\Sigma_1$ is called {\it the dual quark condensate} or {\it dressed Polyakov loop} and transforms, under center symmetry, analogously to the Polyakov loop \cite{pol,suss,svet}, therefore offering a novel order parameter for confinement (see Ref.~\cite{erek} for details).

The procedure of Ref.~\cite{flachitanaka} can be easily adapted to compute the dual fermion condensate for the above boundary conditions (\ref{bcs}). The basic steps are similar to Ref.~\cite{flachitanaka} and we will be briefly repeat them here for the convenience of the reader. As explained in the introduction, we need here to couple the fermion condensate to the Polyakov loop given by an SU($N_c$) matrix in color space,
\bea
L(x)= \mathcal{T} \exp\left[-\imath \int_0^\beta d\tau A_0(\tau, x)\right]~,
\eea
where 
$\mathcal{T}$ is a path-ordering operator and $A_0(\tau, x)$ represents a background, homogeneous, static, temporal component of the gluon field (with the gauge fields quantized according to the usual periodic boundary conditions). The coupling can be realized as in Ref.~\cite{fukushima} where the Polyakov loop is identified with a imaginary quark chemical potential. As in Ref.~\cite{fukushima}, we will neglect any dependence of the coupling constant $\lambda$ on $L$.
The effective action for $\sigma_\varphi = - {\lambda\over N}\langle \bar\psi \psi\rangle_\varphi$ can be expressed, at leading order in a derivative expansion of the Polyakov loop, as 
\bea
\mathscr{S}_{eff}^{(\varphi)} = -\int d^{D}x \sqrt{g} \left({\sigma_\varphi^2\over 2\lambda}\right) 
+{1\over 2} \sum_{\epsilon=\pm 1}  \sum_{n=-\infty}^\infty 
\mbox{Tr} \log \, \mathscr{D}_n^{(\varphi)},\nonumber
\eea
where the trace is over the Dirac, spatial and color indices, and
\bea
\mathscr{D}_n^{(\varphi)}
&=& - \Delta + {1\over 4}R + \mbox{X}_\varphi^2 + \omega_n^2(\varphi) -\left(\mu-\imath \varpi\right)^2\nonumber\\
&& -2 i \left(\mu-\imath \varpi\right) \omega_n(\varphi), \nonumber
\eea
where $\mbox{X}_\varphi^2 = m^2+ \sigma_\varphi^2 + \epsilon \left|\partial \sigma_\varphi\right|$ and the generalized frequencies given by
\bea
\omega_n(\varphi) &=& {2\pi\over \beta}\left(n+{\varphi\over 2\pi}\right)~.\label{oijphi}
\eea
The Polyakov loop is related to the real-valued function $\varpi$ by the relation $\imath \varpi \equiv \ln L(x)$.
The operator $\Delta$ is the Laplacian over the spatial section, $\mu$ the chemical potential (that can take both real and complex values), and $\epsilon = \pm 1$ come from the summation over the eigenvalues of the gamma matrices with the tetrad frame chosen as in \cite{flachitanaka}.

Taking the Mellin transform of the heat trace, we can define the following complex-valued, $\varphi$-dependent zeta function
\bea
\zeta_{\varphi} (s) =  {1\over \Gamma(s)} \sum_{n, \epsilon} \int_0^\infty dt\,
t^{s-1} \mbox{Tr}\,e^{-t \mathscr{D}_n^{(\varphi)}}
\label{zetan}
\eea
and write $\mathscr{S}_{eff}$ as analytical continuation of $\zeta_\varphi(s)$ and its derivative to $s=0$. An appropriate expansion of the propagator may then be used to express the effective action in terms of integrals over geometrical invariants. 

Situations with spatially constant or slowly varying condensates can be accommodated by using the following expansion \cite{tomsparker,Jack:1985mw} (the different situation of rapidly varying condensates will be discussed in the Appendix):
\bea
\mbox{Tr}\,e^{-t \mathscr{D}_n^{(\phi)}} = {1\over (4\pi t)^{d\over 2}} \mbox{Tr}_c \,e^{-t \mathscr{Q}} \sum_k \mathscr{C}^{(k)}_{\epsilon}\, t^k,
\label{ht}
\eea
where $\mbox{Tr}_c$ is the trace in color space and $\mathscr{Q}_{\varphi}= \mbox{X}_\varphi^2 + R/12 + \omega_n^2(\varphi) -\left(\mu-\imath \varpi\right)^2 -2 i \left(\mu-\imath \varpi\right) \omega_n(\varphi)$. Simple steps allows us to obtain the following formula
\bea
\zeta_{\varphi}(s) =  {1\over \Gamma(s)} \sum_{k,\epsilon} 
\int_0^\infty dt\, {t^{s-1+k} \over (4\pi t)^{d\over 2}} 
\mathscr{C}_{\epsilon}^{(k)} e^{-t\mathscr{M}_{\epsilon}(\varphi)} \mathscr{F}_{\beta,\mu}(t),~~~~
\label{zz}
\eea
where $\mathscr{M}_{\epsilon}(\varphi) = R/12 +\mbox{X}_\varphi^2$ and with the thermodynamical kernel given by
\bea
\mathscr{F}_{\beta,\mu,\varpi}(t, \varphi) &=& \sum_{n=-\infty}^\infty e^{-t\left( \omega_n^2(\varphi) -2 i \left(\mu-\imath \varpi\right) \omega_n(\varphi)-\left(\mu-\imath \varpi\right)^2\right)}.\nonumber~~~
\eea
Notice that neither the functional form nor the factorization property of the generalized zeta function change for the above U$(1)$-valued boundary conditions with respect to those used in Ref.~\cite{flachitanaka}. Only the thermodynamical kernel acquires an explicit dependence on the phase, while both in the quantity $\mathscr{M}_{\epsilon}(\varphi)$ and the heat-kernel coefficients $\mathscr{C}^{(0)}_{\epsilon} = 1~, \mathscr{C}^{(1)}_{\epsilon} = 0~, \mathscr{C}^{(2)}_{\epsilon} = \mathscr{R} + {1\over 6} \Delta \left(\sigma_\varphi^2 + \lambda \left|\partial \sigma_\varphi\right|\right)$ with $\mathscr{R}={1\over 180}R_{\mu\nu\rho\sigma} R^{\mu\nu\rho\sigma} -{1\over 180}R_{\mu\nu} R^{\mu\nu}- {1\over 120} \Delta R$, the dependence is only implicit (through the dual condensate) and for the latter disappears in the spatially constant case. Already at this stage, we can anticipate that the same scaling properties of the effective action noticed in Ref.~\cite{flachitanaka} will hold for the present case. 

Using standard series representations for elliptic theta functions allows us to rearrange the thermodynamical kernel as follows 
\bea
\mathscr{F}_{\beta,\mu,\varpi}(t, \varphi) &=& 
{\beta\over 2\sqrt{\pi t}} \left[1 +2 \sum_{n=1}^\infty e^{-\beta^2n^2\over 4t} \Upsilon^{(n)}_{\varphi}(\beta,\mu,\varpi)
\right],
\nonumber
\eea
where
\bea
\Upsilon^{(n)}_{\varphi}(\beta,\mu,\varpi) &=& \chi_n^{(1)}\left(\varphi\right)\cosh\left(\beta \mu n\right)+
+\imath \chi^{(2)}_n\left(\varphi\right) \sinh\left(\beta \mu n\right),\nonumber
\eea
where
\bea
\chi^{(1)}_n\left(\varphi\right)&\equiv&
\cos\left({n\varphi}\right) \cos\left(\beta \varpi n\right) 
+\sin\left({n\varphi}\right) \sin\left(\beta \varpi n\right)\nonumber\\
\chi^{(2)}_n\left(\varphi\right)&\equiv&\cos\left({n\varphi}\right) \sin\left(\beta \varpi n\right) -
\sin\left({n\varphi}\right) \cos\left(\beta \varpi n\right).\nonumber
\eea
The boundary conditions (\ref{bcs}) induce an imaginary part in the effective action, and a straightforward computation gives for the real ($\Re$) and imaginary ($\Im$) parts of the zeta function the following expressions:
\bea
\Re \zeta_\varphi (s) &=&  {\beta\over (4\pi)^{D\over 2}} \sum_{k,\epsilon} \Big(  \mathscr{C}^{(k)}_{\epsilon} \mathbb{I}^{(k)}_{\epsilon, \varphi} + 
\nonumber\\&+& 
\sum_{n=1}^\infty \chi^{(1)}_n\left(\varphi\right)  \cosh(\beta \mu n) \mathscr{C}^{(k)}_{\epsilon} \mathbb{J}^{(k,n)}_{\epsilon, \varphi}\Big)\nonumber
\eea
\bea
\Im \zeta_\varphi (s) &=&  - {\beta\over (4\pi)^{D\over 2}} \sum_{k,\epsilon} \sum_{n=1}^\infty \chi^{(2)}_n\left(\varphi\right)  \sinh(\beta \mu n) \mathscr{C}^{(k)}_{\epsilon} \mathbb{J}^{(k,n)}_{\epsilon, \varphi},\nonumber
\eea
with
\bea
\mathbb{I}^{(k)}_{\epsilon, \varphi} &=& 
{\Gamma(s+k-D/2)\over \Gamma(s)}\mathscr{M}^{D/2-k-s}_{\epsilon}(\varphi)~,\nonumber\\
\mathbb{J}^{(k,n)}_{\epsilon, \varphi} 
&=& 
{2^{D/2+1-k-s}\over \Gamma(s)} \left({\mathscr{M}_{\epsilon}(\varphi)\over n^2\beta^2}\right)^{D/4-(k+s)/2}\times
\nonumber\\&\times& 
K_{k+s-D/2}\left(n\beta\sqrt{\mathscr{M}_{\epsilon}(\varphi)}
\right)~.\nonumber
\eea
Proceeding with the analytical continuation to $s=0$ we arrive at the following expression for the effective action 
\begin{widetext}
\bea
\mathscr{S}_{eff}^{(\varphi)} &=& -\int d^Dx \sqrt{g} {\sigma_\varphi^2\over 2 \lambda}
+ {1 \over 2} {\mbox{Tr}_c \over (4\pi)^{D/2}} \sum_{\epsilon} \int d^dx \sqrt{g} \, \left[
\sum_{k=0}^{[D/2]} \gamma_{k}(D) \mathscr{C}^{(k)}_{\epsilon} 
\mathscr{M}^{D/2-k}_{\epsilon}(\varphi) \ln\left(\ell^2 \mathscr{M}_{\epsilon}(\varphi) \right)
\right.\nonumber\\
&+&
\sum_{k=0}^\infty \left(
a_k(D) \mathscr{C}^{(k)}_{\epsilon}\mathscr{M}^{D/2-k}_{\epsilon}(\varphi)
+2^{D/2+1-k}
\mathscr{C}^{(k)}_{\epsilon}
\mathscr{M}^{D/4-k/2}_{\epsilon}(\varphi)
\left( \Phi^{(+)}-\imath\,\Phi^{(-)}\right) \right],
\label{effact}
\eea
\end{widetext}
where we have introduced the shorthand notation
\bea
\Phi^{(+)} &=& \sum_{n=1}^\infty
\chi^{(1)}_n\left(\varphi\right)
{\cosh(\beta \mu n)
\over
\left(n\beta\right)^{D/2-k}} 
K_{k-D/2}\left(n\beta\sqrt{\mathscr{M}_{\epsilon}(\varphi)}\right)\nonumber\\
\Phi^{(-)} &=&
\sum_{n=1}^\infty
\chi^{(2)}_n\left(\varphi\right)
{\sinh(\beta \mu n)
\over
\left(n\beta\right)^{D/2-k}} 
K_{k-D/2}\left(n\beta\sqrt{\mathscr{M}_{\epsilon}(\varphi)}\right)
\nonumber
\eea
and defined
\bea
{\Gamma(s+k-D/2)\over \Gamma(s)} := \gamma_{k}(D) + s a_k(D) + O(s^2) ~.\nonumber
\eea
Expression (\ref{effact}) gives the effective action for the phase-dependent fermion condensate from which the dual can be computed using the definition (\ref{dual}) after minimization of the effective action. 

The above form for the effective action is not restricted to the case of spatially constant condensates, but it is valid also for the more general situation where spatial variations (in the geometry or in the condensate) are included. We note here that the heat trace expansion (\ref{ht}) used to obtain (\ref{effact}) still contains a gradientlike expansion and implicitly assumes that variations of the condensate, of the geometry (or of any external field one may wish to add) do not occur rapidly over space. Physically, quantities may be large in magnitude, but the derivatives should be small. The rapidly varying case can be analyzed using a different form for the heat-trace and we will show how to do this in the Appendix. 
For the case of spacetimes with constant curvature and in the approximation of spatially constant condensates, the above formula considerably simplifies.

The reader may check that several of the expected properties of the effective action are clearly encoded in the above representation. Nicely enough, the same scaling property of the effective action, discussed in Ref.~\cite{flachitanaka} for the chiral condensate, holds for the dual one. Basically, once $\chi \rightarrow \kappa \chi$ where $\chi$ is any quantity with mass dimensions ($\chi=m,~\mu,~\lambda^{-1/2},~\sigma_\varphi,~\beta^{-1},~\varpi,~\mu,~\ell^{-1}$), the effective action rescales as $\mathscr{S}_{eff} \rightarrow \kappa^{4} \mathscr{S}_{eff}$ allowing one to fix one of the parameters arbitrarily. Also, the Roberge-Weiss periodicity \cite{roberge} (see also Ref.~\cite{sakai}) is evident once the chemical potential is transformed as $\mu \rightarrow \imath \mu/T$. 

One other advantage of the present approach is that it allows a direct computation of the imaginary part of the effective action that appears due to the U$(1)$-valued boundary conditions. In fact, within a a zeroth order approximation that treats the imaginary part as perturbation, it is usually argued, for example in Refs.~\cite{ruggieri,mukherjee,sasagawa}, that this quantity is small, causing negligible changes in the dual condensate (see Ref.~\cite{sasa2} for a quantitative justification of this point). Here, we will follow the above examples and verify, in the numerical analysis of Sec.~\ref{sec4}, that the imaginary part remains negligibly small for the range of parameters considered.

We should remark here that the full effective action should include also the purely gluonic part, {\it i.e.},
\bea
\mathscr{S}_{tot}\left(\varpi, \sigma_{\varphi}\right) =  \mathscr{S}_{glue}\left(\varpi\right) + \mathscr{S}^{(\varphi)}_{eff}\left(\varpi,\sigma_{\varphi}\right)~.
\eea
In flat space, an expression for $\mathscr{S}_{glue}\left(\varpi\right)$ can be obtained using the leading order strong coupling result that depends on some free parameter, finally fitted against the lattice data (see \cite{fukushima}). In curved space an analogous procedure is currently not possible both for the lack of any explicit expression for $\mathscr{S}_{glue}\left(\varpi\right)$ and of any numerical result. Despite this fact, the dual fermion condensate can be computed as a function of the Polyakov loop, for which one may use, in specific cases, phenomenological parametrizations. 

\section{Density of States Method}
\label{sec3}
In the present section we will discuss yet another approach to compute the effective action based on the density of states method. This proves to be useful when the chemical potential becomes large. In fact, for $\Re \mu \neq 0$, the condition $\Re \left(\mu - \mathscr{M}_\epsilon(\varphi)\right)<0$, met for large enough values of the curvature $R$ or of the bare mass $m$ with respect to the chemical potential, has to hold to ensure the convergence of the sum over the modified frequencies and, in turn, of the representation (\ref{effact}). As before, we may follow Ref.~\cite{flachitanaka} modifying the procedure described there to include the U$(1)$-valued boundary conditions (\ref{bcs}). We start from
\bea
\zeta'_\varphi(0) 
= {1\over (4\pi)^{d/2}} \sum_{k,\epsilon} \mathscr{C}_\lambda^{(k)} \mathscr{Z}_\epsilon^{(k)}~,
\label{Z}
\eea 
with
\bea
\mathscr{Z}_\epsilon^{(k)} &=& 
\lim_{s\rightarrow 0}{d\over ds}
{1\over \Gamma(s)} 
\int_0^\infty dt\, t^{s-1+k - d/2-1/2} 
e^{-t\mathscr{M}_{\epsilon}(\varphi)} 
\times\nonumber\\
&\times&{\beta\over \sqrt{\pi}}
\left[
1+ \sum_{n=-\infty}^{~~\infty~~\prime} e^{- \imath n\varphi} e^{-\beta^2 n^2\over 4 t}e^{-\beta\mu n -\imath \varpi n}
\right].
\label{zet}
\eea
Defining the phase-dependent density of states $\rho_\epsilon^{(k)}(E,\varphi)$ as
\bea
{e^{-t \mathscr{M}_\epsilon(\varphi)}\over t^{d/2-k}} := {1\over 2} \int_0^{\infty} dE E \rho_\epsilon^{(k)}(E,\varphi) e^{-t E^2}~,
\label{density}
\eea
expression (\ref{Z}) can be written as
\bea
\mathscr{Z}_\epsilon^{(k)} = 2 \int_0^\infty dE E \rho_\epsilon^{(k)}(E,\varphi) I(E) + \mathscr{Z}_{0},
\label{I}
\eea
where 
\bea
I(E) &=& {\beta\over 2\sqrt{\pi}}
\sum_{n=-\infty}^{~~\infty~~\prime} e^{- \imath n\varphi}
\int_0^\infty {dt\over \sqrt{t}} \times\nonumber\\
&\times& \int_E^\infty dx x e^{-t x^2} e^{-\beta^2 n^2\over 4 t} e^{-\beta\mu n-\imath \varpi n}.\label{Izt}
\eea
The first term in the right hand side of Eq.~(\ref{I}) corresponds to the sum over $n$ in (\ref{zet}) (the second term in square brackets), while the second in the right hand side of Eq.~(\ref{I}) term comes from the $n=0$ contribution in (\ref{zet}) (the first term in square brackets). Using the following identity:
\bea
{\beta\over 2\sqrt{\pi}} e^{-\imath n\varphi} {e^{{-\beta^2 n^2\over 4 t} - \beta\mu n-\imath \varpi n}\over \sqrt{t}}
=
\int_{-\infty}^{+\infty} {dz\over 2\pi}\, e^{\imath n z} e^{-t\left({1\over \beta}(z+\varphi+\varpi)-\imath \mu\right)^2},
\nonumber
\eea
we may write (\ref{Izt}) as
\bea
I(E) &=&
{1\over 2\pi} 
\sum_{n=-\infty}^{~~\infty~~\prime} 
\int_0^\infty {dt} \int_E^\infty dx x e^{-t x^2}\times\nonumber\\
&\times& \int_{-\infty}^{+\infty} dz\,
e^{\imath n z} e^{-t\left({1\over \beta}(z+\varphi+\varpi)-\imath \mu\right)^2}.
\nonumber
\eea
\begin{widetext}
Integrating, in turn, over $t$, $z$, summing over $n$, and finally integrating over $x$ leads us to the following expression
\bea
I(E) &=&I_\varphi^{(0)}(E) -\ln \left({1 + \varepsilon_\varphi L e^{-\beta (E+\mu)}\over 1 + \varepsilon_\varphi L e^{-\beta E}} \right) 
-\ln \left({1 +\varepsilon_\varphi^* L^* e^{-\beta (E-\mu)}\over 1 + \varepsilon_\varphi^* L^* e^{-\beta E}} \right)
\label{th}
\eea
where $\varepsilon_\varphi=\exp\left({\imath (\varphi-\pi)}\right)$. The quantity $I_\varphi^{(0)}(E)$ combined with with $\mathscr{Z}_0$ gives the contribution for vanishing $\mu$ that can be expressed as in (\ref{effact}) with $\mu=0$. The final form of the effective action is then found to be
\bea
\mathscr{S}_{eff}^{(\varphi)} = \mathscr{S}_{eff}^{(\varphi)}\Big|_{\mu=0} -{1\over (4\pi)^{d/2}} \int d^d x \sqrt{g} \int_0^\infty dE E \rho(E) 
\left[
\mbox{Tr}_c\,\ln \left({1 + \varepsilon_\varphi L e^{-\beta (E+\mu)}\over 1 + \varepsilon_\varphi L e^{-\beta E}} \right)
+\mbox{Tr}_c\,\ln \left({1 +\varepsilon_\varphi^* L^* e^{-\beta (E-\mu)}\over 1 + \varepsilon_\varphi^* L^* e^{-\beta E}} \right)
\right],
\label{seffmu}
\eea
\end{widetext}
where we have defined the function
\bea
\rho(E,\varphi) := \sum_{k} \mathscr{C}^{(k)}_\epsilon \rho_\epsilon^{(k)}(E,\varphi).
\eea
The result of Ref.~\cite{flachitanaka} for the chiral condensate is recovered by setting $\varphi=\pi$. 

The difficulty in using the approach presented in this section lies in computing the density of states. There are several ways to proceed and here we will describe two such possibilities. A direct way is to follow the same strategy outlined Ref.~\cite{flachitanaka} that applies to the present case with no modifications. Starting from the definition of the density of states (\ref{density}), it is easy to obtain the following recursive relation
\bea
\rho_\epsilon^{(k)}(E,\varphi) &=& 
{\partial^k\over \partial(E^2)^k}\rho_\epsilon^{(0)}(E,\varphi)~,
\eea
that, used along with
\bea
\rho_\epsilon^{(0)}(E,\varphi)&=& 4 \pi^{-d/2} \int d^d p\, \delta\left(E^2 -\mathscr{M}_\varepsilon(\varphi)-p^2\right)~,\nonumber
\eea
allows to perform the integration over $E$ in (\ref{seffmu}) leaving the integration over $p$ that can be done numerically.

Although the above method may be preferred in numerical evaluations, here we wish to describe another approach based on the use of Tauberian theorems \cite{wiener,baltes,tomsdensity,elizalde}\footnote{Ref.~\cite{wiener} presents a thorough discussion of Tauberian theorems; Ref.~\cite{baltes} gives a readable account and several physical applications; Ref.~\cite{tomsdensity} gives some general discussion and an application to the case of harmonic oscillator potentials in the context of Bose-Einstein condensation; Ref.~\cite{elizalde} uses Tauberian theorems as compatibility check in the context of zeta regularization in non-compact domains.}.
Defining 
\bea
{1\over 4}{\mathcal K}_\epsilon^{(k)}(t) := {e^{-t \mathscr{M}_\epsilon(\varphi)}\over t^{d/2-k}} ~,
\label{kcal}
\eea
we may write
\bea
{\mathcal K}_\epsilon^{(k)}(t) = \lim_{h\rightarrow 0^+} \int_h^{\infty} dE^2 \rho_\epsilon^{(k)}(E,\varphi) e^{-t E^2}~,
\label{tb}
\eea
that is recognized as the Laplace transform of the density of states. 
Assume that, for $0 < t < q \in \mathbb{R}$,
\bea
{\mathcal K}_\epsilon^{(k)}(t) = \sum_{i=1}^{i_{max}} c_i t^{-r_i} + O\left(
t^{-r_{i_{max}}+1}
\right),
\label{sK}
\eea
with $0 < r_i<r_{i+1}$ and $i\in \mathbb{N}/{0}$. For ${\mathcal K}_\epsilon^{(k)}(t)$ defined in (\ref{kcal}) one finds
\bea
r_i &=& {d\over 2} -k - i,\\
c_i &=& {(-1)^i \over i!} \mathscr{M}^i_{\epsilon}(\varphi).
\eea
For $r_i>0$, it is enough to use the definition of the Gamma function,
\bea
t^{-r_i} = {1\over \Gamma(r_i)} \int_0^{\infty} dE^2 \left(E^2\right)^{r_i-1} e^{-t E^2},
\eea
along with relations (\ref{tb}) and (\ref{sK}) to arrive at
\bea
\rho_\epsilon^{(k)}(E) \simeq  \sum_{i=1}^{i_{max}} {c_i \over \Gamma(r_i)} \left(E^2\right)^{r_i-1}.
\eea
Unfortunately, the condition $r_i>0$ is only met for $k+i<d/2$ (for $d=3$ and for any positive $i$ and non-negative $k$ it never occurs). This limitation can be overcome by using Tauberian theorems (see page 30-31 of Ref.~\cite{baltes} and references given there). The basic result can be stated as follows. Consider (\ref{tb}) and assume that $\rho_\epsilon^{(k)}(E,\varphi)$ is a smooth function. Assume that for $0 < t \leq t_{*}$ relation (\ref{sK}) holds with the exponents $r_i \in \mathbb{R}$ for any $i$. Under these assumptions (the reader can easily verify these to be satisfied in the present case), then for $E^2 > h$,
\bea
\rho_\epsilon^{(k)}(E,\varphi) &\simeq&  \sum_{i=1}^{i_{max}} {c_i \over \tilde{\Gamma}(r_i)} \left(E^2\right)^{r_i-1}+\nonumber\\
&+& O\left((E^2)^{r_{i_{max}+1}}\ln \left((E^2)\right)
\right).\label{karamatas}
\eea
In the above formula $\tilde{\Gamma}(r_i)$ is defined to be zero for any nonpositive integer, while $\tilde{\Gamma}(r_i)=\Gamma(r_i)$ for any other value of the argument. Substituting (\ref{karamatas}) in (\ref{seffmu}) makes the $E$ dependence explicit allowing straightforward integration. The integration can be performed analytically in some cases (even $d$) or, more generally, by numerical approximation. However, due to the condition $E^2 > h$, some care has to be used with the IR part of the integration range, {\it i.e.}, with the limit $h \rightarrow 0$ in (\ref{tb}) for negative $r_i$. In this case, the integration can be performed by splitting the integration range, $\left(0, \infty\right) = \left(0, \xi\right] \cup \left(\xi, \infty\right)$ with $\xi$ small. Integrating over $\left(\xi, \infty\right)$ poses no problem, while the integral over $\left(0, \xi \right]$ can be performed by keeping the dimensionality general and proceeding by analytical continuation.

\section{Constant Curvature Case}
\label{sec4}

In this section we will look in some detail, as a test of the formal results discussed in the preceding sections, at the case of constant curvature spacetimes. This will serve, first of all, as a concrete application, similar to that discussed in Ref.~\cite{sasagawa}. While in the case of spatially constant condensates we can recover, using a different approach, the main features discussed in Ref.~\cite{sasagawa}, our main goal is to extend the analysis and include an explicit coupling to the Polyakov loop as well as inhomogeneous phases, since such a procedure is essential to study what happens in the vicinity of a black hole (we will discuss more about this point in the concluding section). Compared to the case discussed in Ref.~\cite{flachitanaka} where chiral symmetry breaking in curved space was studied, here the problem presents additional computational needs due to the explicit dependence of the effective action on the phase $\varphi$.

Practically, we shall consider the most representative geometry of the above type, that is the Einstein universe, which is topologically $\mathbb{R}\otimes \mathbb{S}^3$. Explicitly, the metric takes the form
\bea
ds^2 = dt^2 - a^2 \left( d\vartheta^2 + \sin^2\vartheta \left(d\phi^2 + \sin^2 \phi^2 d\chi^2\right)\right),~
\eea
where $0 \leq \vartheta \leq \pi$, $0 \leq \phi \leq \pi$, $0 \leq \chi \leq 2\pi$.
The Ricci scalar is related to the curvature radius of the sphere, $a$, by $R=6/a^{2}$. The numerical analysis of this section will be specialized to this case. With respect to the analogous computation for the chiral condensate, the numerical procedure requires some modification
in the summations over the modified frequencies due to the dependence on the phase $\varphi$ caused by the U$(1)$-valued boundary conditions. Contrary to the case of Ref.~\cite{flachitanaka}, for the inhomogeneous case the explicit $\varphi$ dependence will make the problem effectively two dimensional requiring some additional computational effort. The Fourier transform (\ref{dual}) is performed using standard numerical techniques. For simplicity, we will perform the analysis of this section for $d=3$ and limit ourselves to the case of small chemical potentials for which we may use the formulation described in Sec.~\ref{sec2}.

The analysis is performed first for the case of spatially constant condensates for which we only need to minimize the effective potential obtained from (\ref{effact}). We will do so by discretizing the $\varphi$ direction, computing the minima for fixed $\varphi$ and finally computing the Fourier transform numerically.  
Before minimization, the thermodynamic potential is normalized by subtracting its value for $\sigma_\varphi=0$. Owing to the scaling properties discussed at the end of Sec.~\ref{sec2}, we will fix the renormalization scale to $\ell$ as indicated and measure all quantities accordingly.

The simplest way to fix the dependence on the Polyakov loop is to use a simple parametrization, essentially modeling its dependence from the temperature as a step function with the position of the step taken as a free parameter. In fact, what we do here is to take the quantity $\varpi$ at fixed temperature as a free parameter that we change within the interval $\varpi=0, \sigma_0/2, \sigma_0$ ($\sigma_0$ indicates the value of the chiral condensate at $T=0$) to have an idea of how much the dual condensate can be affected by a proper inclusion of gauge degrees of freedom and take the difference $\delta \varepsilon = (\sigma_\varphi (\sigma_0)-\sigma_\varphi (0))/2$ as an estimate for this error. This is certainly an approximate way to proceed, that we accept in exchange for being able to introduce effects of curved space. 

\begin{figure}[ht]
\begin{center}
\unitlength=1mm
\begin{picture}(155,30)
   \includegraphics[height=3.1cm]{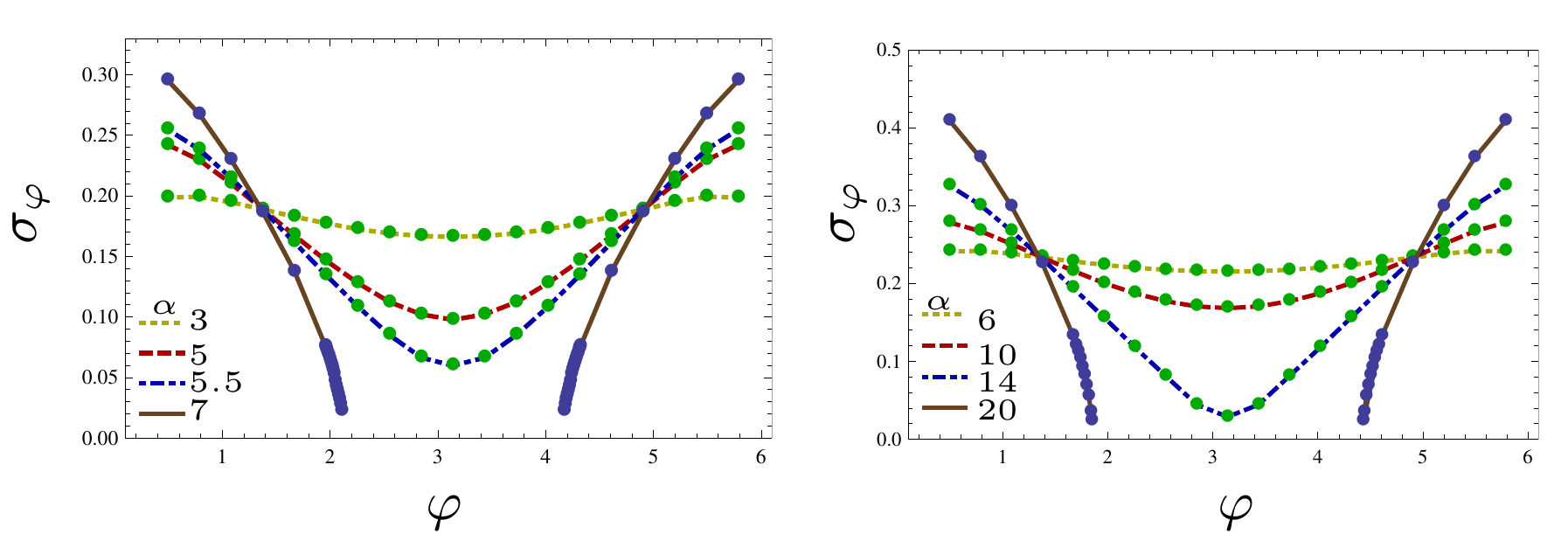}
\end{picture}
\end{center}
\caption{
Left: Numerical solution for the quantity $\sigma_\varphi$ {\it vs} $\varphi$ for $\mu=0$ and $a=10$, $m=0$ and $\varpi=0$. The dots are computed numerically while the continuous curves are obtained by interpolation. Right: Numerical solution for the quantity $\sigma_\varphi$ {\it vs} $\varphi$ for $\mu=0.1$ and $a=20$, $m=0$ and $\varpi=0$. The quantity $\alpha=a\times T$.} 
\label{fig1}
\begin{center}
\unitlength=1mm
\begin{picture}(155,30)
   \includegraphics[height=3.1cm]{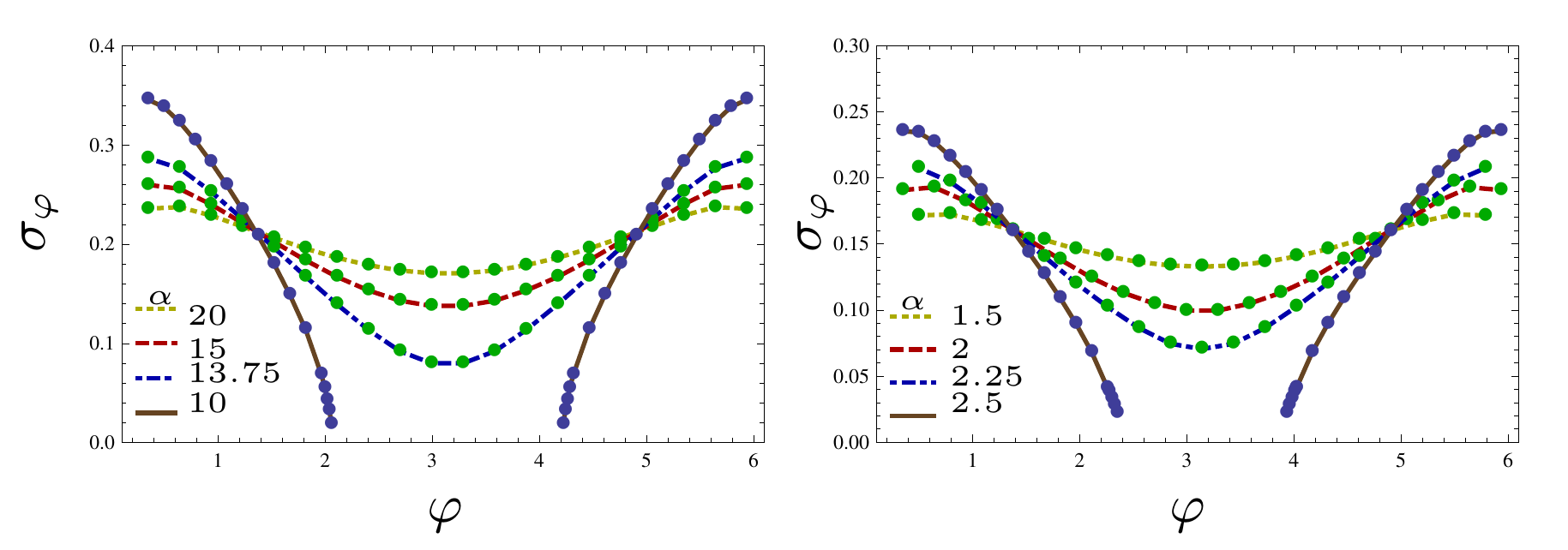}
\end{picture}
\end{center}
\caption{
Numerical solution for the quantity $\sigma_\varphi$ {\it vs} $\varphi$ with the dependence on the Polyakov loop included. This dependence has been fixed by parametrizing the Polyakov loop as a constant that we change within the interval $\varpi=0, \sigma_0/2, \sigma_0$. The central value refers to $\sigma_\varphi (\varpi=\sigma_0/2)$ while the error is estimated by $\delta \varepsilon = (\sigma_\varphi (\varpi=\sigma_0)-\sigma_\varphi (\varpi=0))/2$. The left-hand (right-hand) panel refers to $\mu=0$, $a=20$ and $m=0$ ($\mu=0.1$, $a=10$ and $m=0.1$). The quantity $\alpha=a\times T$.} 
\label{fig2}
\end{figure}

\begin{figure}[h]
\begin{center}
\unitlength=1mm
\begin{picture}(175,30)
   \includegraphics[height=3.1cm]{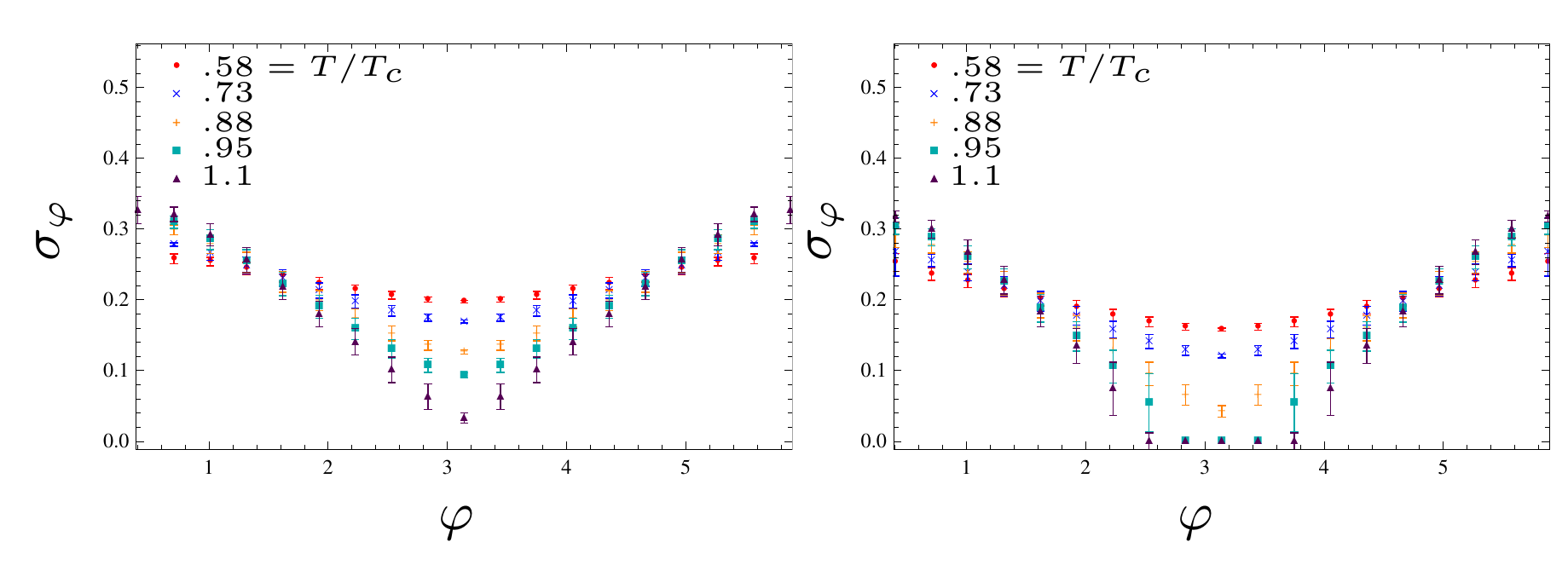}
\end{picture}
\end{center}
\caption{For sample sets of parameter how the dual condensate depends on the Polyakov loop. The central value indicates $\sigma_\varphi (\sigma_0/2)$ and the error bars are computed according to $\delta \varepsilon = (\sigma_\varphi (\sigma_0)-\sigma_\varphi (0))/2$. The lack of accuracy due to ignoring the Polyakov loop is $O(10\%)$. } 
\label{fig3}
\end{figure}

\begin{figure}[ht]
\begin{center}
\unitlength=1mm
\begin{picture}(100,30)
   \includegraphics[height=3.cm]{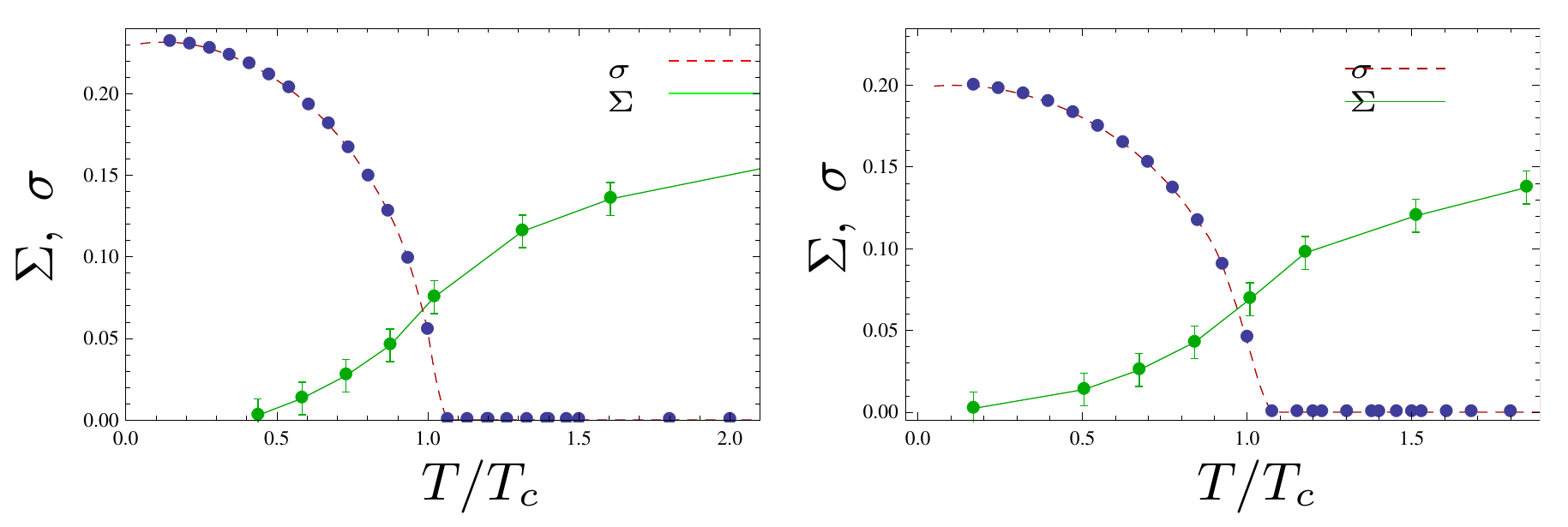}
\end{picture}
\end{center}
\caption{Left: Chiral (dashed red line) and dual (continuous green line) condensates for $\mu=0$, $m=0$, $a=20$. The critical temperatures are $T_c = 0.68$ for the fermion condensate and $T_{c} = 0.72$ for the dual. Right: Chiral (dashed red line) and dual (continuous green line) condensates for $\mu=0.1$, $m=0.1$, $a=10$. The critical temperatures are $T_c = 0.60$ for the fermion condensate and $T_{c} = 0.66$ for the dual.} 
\label{fig4}
\end{figure}

\begin{figure}[ht]
\begin{center}
\unitlength=1mm
\begin{picture}(150,30)
   \includegraphics[height=3.2cm]{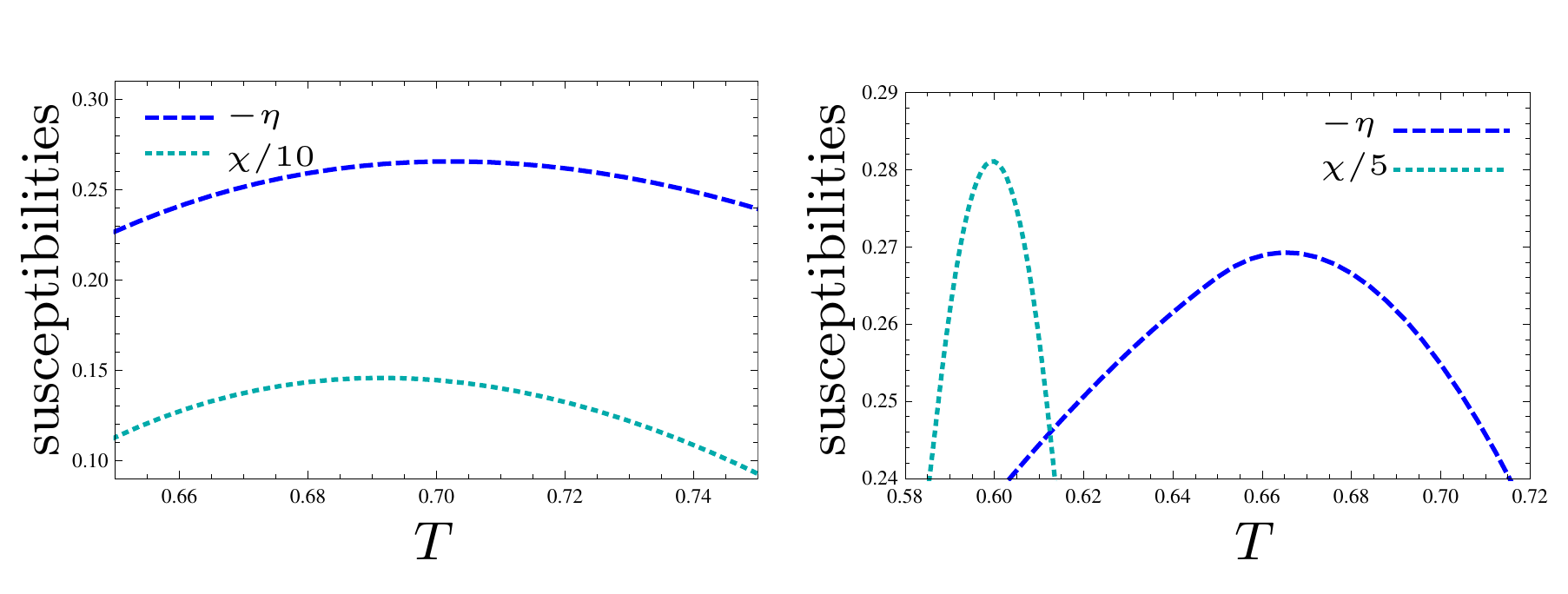}
\end{picture}
\end{center}
\caption{Susceptibilities around the peaks for smaller and larger values of the curvature. Left: $\mu=0$, $m=0$, $a=20$. Right: $\mu=0.1$, $m=0.1$, $a=10$.} 
\label{fig5}
\end{figure}

\begin{figure}[ht]
\begin{center}
\unitlength=1mm
\begin{picture}(1000,60)
   \includegraphics[height=5.9cm]{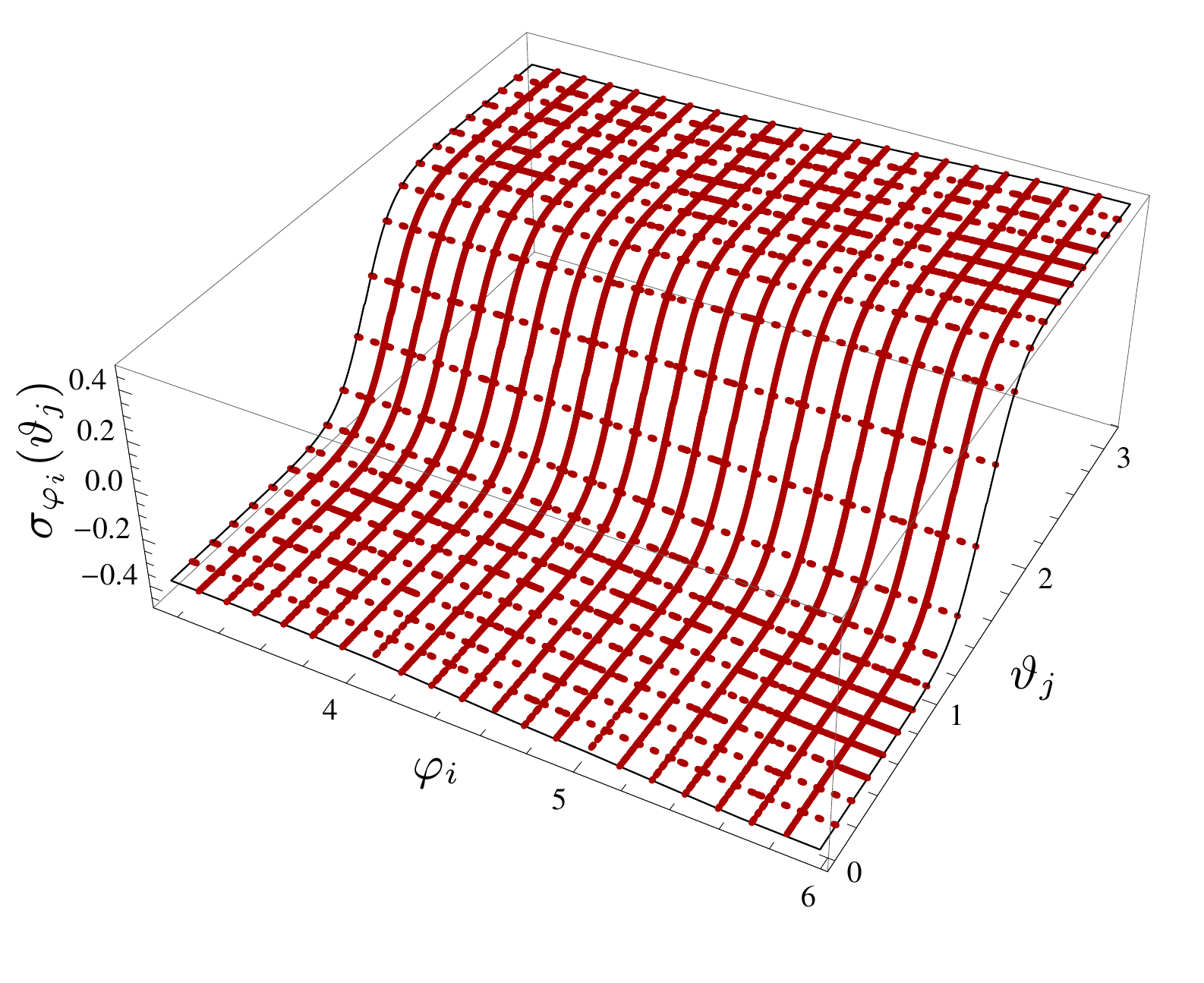}
\end{picture}
\end{center}
\caption{Full three-dimensional dual condensate $\sigma_{\varphi_i}(\vartheta_j)$. The parameters are set to $\lambda=10$, $\ell^{-1}=10^3$, $T=0.47$, $a=10$.} 
\label{fig6}
\end{figure}

\begin{figure}[ht]
\begin{center}
\unitlength=1mm
\begin{picture}(150,30)
   \includegraphics[height=3.2cm]{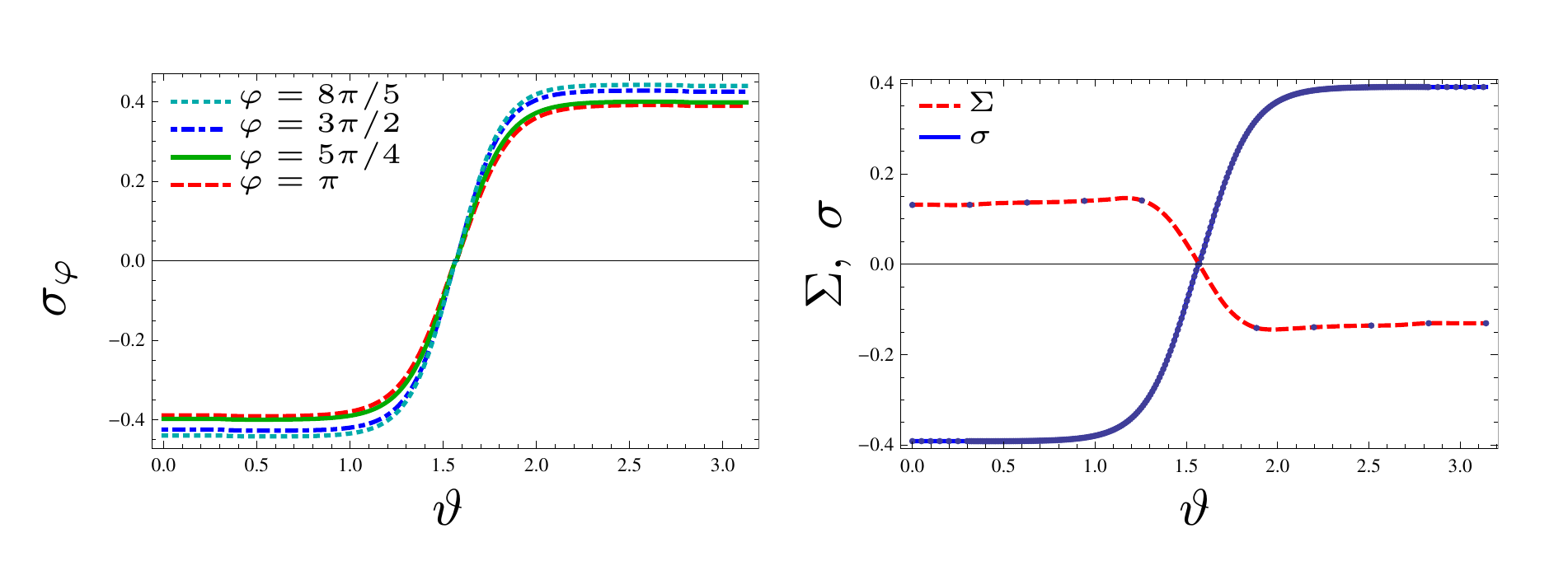}
\end{picture}
\end{center}
\caption{Left: Spatial dependence of the dual condensate for fixed phase $\varphi$. Right: Dual ($\Sigma\equiv \Sigma_1$) {\it vs} chiral ($\sigma\equiv \sigma_\pi$) condensate. In both panels the parameters are fixed as in Fig.~\ref{fig5}. 
} 
\label{fig7}
\end{figure}

Our results are illustrated in Figs.~\ref{fig1}--\ref{fig7}. Figures \ref{fig1} and \ref{fig2} show the dependence of $\sigma_\varphi$ on the phase $\varphi$ for a sample set of parameters and for $\varpi=0$. Fig.~\ref{fig3} shows how the dual condensate depends on the Polyakov loop also for some indicative values of the parameters. The central value of each point represents the value $\sigma_\varphi (\varpi=\sigma_0/2)$ while error bars are computed according to $\delta \varepsilon = (\sigma_\varphi (\varpi=\sigma_0)-\sigma_\varphi (\varpi=0))/2$. The lack of accuracy due to ignoring the Polyakov loop is of order $O(10\% - 20\%)$. The dual and chiral condensates, $\sigma\equiv \sigma_\pi$ and $\Sigma \equiv \Sigma_1$, are plotted in Fig.~\ref{fig4} for the same parameters used in Fig.~\ref{fig3} where we have approximated the error due to the Polyakov loop to be exactly $10\%$. Finally, Fig.~\ref{fig5} plots the peaks of the susceptibilities for $\varpi=0$ (to be understood with an error of $O(10 \%)$)

The susceptibilities are defined, respectively for the chiral and dual condensates, by
\bea
\chi &=& {\partial \sigma_\pi \over \partial T}, \nonumber\\
\eta &=& {\partial \Sigma_1 \over \partial T}, \nonumber
\eea
from which the critical temperatures can be computed. Numerically we proceed by fitting the curves obtained for the chiral and dual condensate and then use a numerical maximization routine to find the value for the critical temperatures. Figure \ref{fig4} shows the susceptibilities around the peaks for sample values of the parameters. Similarly to Ref.~\cite{sasagawa}, an increase in the distance between the peaks of the susceptibilities $\chi$ and $\eta$ when curvature increases is also observed here.

We will now extend the analysis to include the spatially varying case. For this case, we will limit our analysis to the case $\varpi=0$. Analogous situations for the chiral condensate have been addressed in Refs.~\cite{flachitanaka,flachirapid,flachiconifold}. To keep the computation as simple as possible, in the following we will set both the mass and the chemical potential to zero and assume the condensate to depend only on the angular direction $\vartheta$. As for the case studied in Ref.~\cite{flachitanaka}, for vanishing $\mu$, we expect inhomogeneous configurations to be metastable (with larger free energy with respect to the homogeneous configuration obtained above), thus characterizing excited states only. The analysis is obviously complicated by the fact that we need to minimize the effective action. The way we carry out the computation is first to discretize the $\varphi$ direction and minimizing the effective action at each point of the $\varphi$ grid. This gives the solution $\sigma_{i}(\vartheta)\equiv\sigma(\varphi_i, \vartheta)$, where $\varphi_i$ represents a generic point of the $\varphi$ grid. The sums over the modified frequencies (\ref{oijphi}) are performed, as in \cite{flachitanaka}, by taking advantage of the exponential decay of the Bessel functions in (\ref{effact}) for a large argument that allow a consistent truncation of the series. Instead, when the argument is smaller than a certain tolerance value, we perform the summation analytically after appropriately expanding the Bessel functions. The full solution is then constructed by matching those obtained in the two regions. We then proceed by discretizing the $\vartheta$ direction for each $\varphi_i$ and compute the Fourier transform numerically. Figures~\ref{fig5} and \ref{fig6} show some sample plots. 
The $\varphi$ dependence of the function $\varphi_{\varphi_i}\left(\vartheta\right)$ is illustrated in the left-hand panel of Fig.~\ref{fig5} for some values of the angle $\varphi$ and of the parameters. The full function $\sigma_{\varphi_i}(\vartheta_j)$ is shown in Fig.~\ref{fig6} for $\pi\leq \varphi\leq 2\pi$ (in the range $0 \leq \varphi\leq \pi$ the solution can be constructed using the reflection symmetry $\varphi_{\varphi}\left(\vartheta\right) = \varphi_{2\pi-\varphi}\left(\vartheta\right)$). Finally, the right-hand panel of Fig.~\ref{fig6} illustrates the spatially varying dual condensate $\Sigma$ plotted together with the chiral condensate $\sigma$.

\section{Discussion}
\label{sec5}

Strongly interacting fermion effective field theories are valuable tools to study chiral symmetry breaking in QCD under the influence of external conditions, may these be due to external sources, finite density, or gravity. These, in fact, greatly complicate the nonperturbative dynamics of QCD and make lattice computations challenging, to say the least. Among the various low energy approximations of QCD, the Nambu-Jona Lasinio model and its variants have been providing for a long time a concrete framework allowing to explore several aspects of QCD, whose description was, otherwise, not feasible (or, in some cases, not well understood) by lattice methods. 
Because of the `fermionic nature' of the Nambu-Jona Lasinio model, the one aspect that remained outside of its realm was an effective way to describe confinement. 
This problem can possibly be amended in several ways, for instance extending the model by including the Polyakov loop \cite{fukushima} or by introducing the dual fermion condensate (see, for example, \cite{Kashiwa:2009ki,ruggieri,mukherjee,Xu:2011pz}). In the present paper we have adopted the latter approach, as described in Refs.~\cite{gattringer,erek} where an order parameter for confinement was engineered directly from the fermion condensate by taking its dual, {\it i.e.} Fourier transforming the condensate computed with respect to a set of U$(1)$-valued boundary conditions (with gauge degrees of freedom respecting periodic boundary conditions). This quantity, as explicitly shown in Refs.~\cite{gattringer,erek},
defines an equivalence class of Polyakov loops with winding number $n$ conjugate to the phase (see the definition (\ref{dual})). Then the dual condensate transforms like the Polyakov loop under center symmetry and therefore can be used as an order parameter to describe the transition to deconfinement. 

In the present paper we have considered a system of strongly interacting fermions in curved space with the fermion condensate coupled to the Polyakov loop. The main point of this work was to translate the formalism of Ref.~\cite{flachitanaka} (designed to describe chiral symmetry breaking in curved space) to the case of the dual fermion condensate and therefore offer a way to construct an order parameter for confinement when the background manifold is curved. Specifically, we have shown how to compute the effective action for the dual fermion condensate as a function of the Polyakov loop at finite temperature and density and extended the procedure of Ref.~\cite{flachitanaka} to include U$(1)$ boundary conditions necessary to construct the dual condensate. We have described two different approaches: one based on the use of heat-kernel methods (specialized to situations in which the geometry and the condensates are slowly or rapidly varying functions in space), and one based on the density of states method. We have then tested the formal results for the case of a constant curvature geometry and constructed, by means of numerical approximation, the dual condensate for both spatially constant and inhomogeneous configurations. 

There are few points, left for future work, deserving a more attentive analysis. A first technical one concerns a deeper analysis of the `twisted' boundary conditions (\ref{bcs}), since these induce an imaginary part in the effective action. For the case of a constant curvature space, studied numerically in Sec.~\ref{sec4}, we have verified that the imaginary part remains negligibly small, therefore providing an {\it a posteriori} justification of the approximation (used here as well as in Refs.~\cite{Kashiwa:2009ki,ruggieri,mukherjee,Xu:2011pz,sasagawa}) that treats the imaginary part as a small perturbation. It is certainly an interesting question to address whether this is true for more general geometries especially when singularities are present and what is the interplay between curvature and, for example, an external electromagnetic field.

The second, most important, point concerns a full inclusion of gauge degrees of freedom. What we have done here is compute the effective action for the dual condensate as a function of the Polyakov loop that is left general. Minimization of the effective action with respect to the dual condensate does not depend directly on the purely gluonic part of the potential, but only indirectly through the Polyakov loop. Clearly, this way of proceeding is imperfect and a full computation of the Polyakov loop potential at strong coupling in curved space is a necessary step to close the story. Here, we have accepted this limitation (in exchange for having a way to explicitly introduce background gravitational effects) and compensated the lack of knowledge of the full gauge field dynamics by supplementing a simple parametrization for the Polyakov loop. A possible alternative way to compute the Polyakov loop potential may be the method of Ref.~\cite{paw} appropriately generalized to curved space.

Generalizing the results of this paper to static geometries should also be possible along the ways outlined in Refs.~\cite{dowker0,dowker1,dowker2}. This becomes particularly relevant if one wishes to describe, within the framework of strongly interacting fermion effective field theories, eventual transitions to deconfinement in the vicinity of a black hole. Providing the basis to address this problem is what motivated us, in the first place, to construct the formalism described in the present work and to test it in the simpler case of a constant curvature geometry. Reasons to approach this problem are various and essentially related to the desire to understand how the radiation of composite degrees of freedom from a black hole occurs, {\it i.e.}, whether composite degrees of freedom are directly emitted, or elementary ones are emitted and composite ones form farther away from the horizon by nonperturbative processes. The problem is clearly complicated, but generalizing the results of this paper to black holes may provide a pathway to, at least, a partial answer. Technically, a way to extend the computations presented here to a black hole geometry can be carried out using the general technique of Refs.~\cite{dowker0,dowker1,dowker2} that consists in transforming the original static (black hole) geometry into an ultrastatic one. Then the effective action can be computed in this related spacetime by adapting the method described in the previous sections. The result in the original spacetime can then be obtained by adding a compensating term, called cocycle function. The whole should, of course, be done for phase dependent boundary conditions necessary to construct the dual condensate. With respect to geometries with constant curvature where the ground state (at least for small chemical potentials and in the absence of boundaries) will stay spatially constant, in the case of black holes both the chiral and the dual condensate will be inhomogeneous, complicating also the numerical analysis involved.

\acknowledgments
The support of the Funda\c{c}\~{a}o para a  Ci\^{e}ncia e a Tecnologia of Portugal and of the Marie Curie Action COFUND of the European Union Seventh Framework Program (Grant Agreement No. PCOFUND-GA-2009-246542) is acknowledged. Thanks are extended to T. Tanaka for discussions and to Y. Gusev for correspondence. I would like to thank K. Fukushima and an anonymous referee for pointing out the importance of adding the Polyakov loop in the computation.

\appendix

\section{Rapidly varying case}
\label{app}

For completeness, in this Appendix we will describe how to deal with situations where the condensate is a rapidly varying function. This case can be dealt with by using a different nonperturbative ansatz for the heat trace, and one possibility is to use covariant perturbation theory developed in Ref.~\cite{BV}. The similar problem for noninteracting fermions at finite temperature and zero chemical potential has been discussed in Ref.~\cite{gusev} that we will follow closely. The present calculation also generalizes the procedure of Ref.~\cite{gusev} to include finite nonvanishing chemical potentials. Setting $\varphi=\pi$ also provides a generalization of the procedure discussed in Ref.~\cite{flachitanaka} to the case rapidly varying chiral condensates.

In the present case, the zeta function (per fermion degree of freedom) takes the same factorized form as before, 
\bea
\zeta_{\varphi} (s) =  {1\over \Gamma(s)} \sum_{\epsilon=\pm 1} \int_0^\infty dt\,
t^{s-1} \mathscr{F}_{\beta,\mu,\varpi}(t, \varphi) \,\mathscr{Z}(t)~
\label{zetan2}
\eea
with the heat trace is written as (in the following the operator $\Box$ should be understood as the $d$-dimensional Laplacian)
\bea
\mathscr{Z}(t) = {1\over (4\pi t)^{d/2}}\left( 1 + t \mathscr{A}_\epsilon + t^2 \mathscr{B}_\epsilon  + \cdots \right),
\label{zbv}
\eea
where
\bea
\mathscr{A}_\epsilon &=& {1\over 4}R + X_\varphi \nonumber\\
\mathscr{B}_\epsilon &=& 
\mathscr{A}_\epsilon \mathscr{F}_1(-t \Box) R +\mathscr{A}_\epsilon \mathscr{F}_2(-t \Box) \mathscr{A}_\epsilon + \cdots. \nonumber 
\eea
In the above formulas the dots represent higher order terms in any of the background fields or curvatures as well as terms that do not depend on the condensate and that disappear upon derivation in the equation of motion. The quantity $R$ represents the Ricci scalar. Using the following integral representations:
\bea
f_1(-t\Box) &=& \int_0^1 du e^{u(1-u)t \Box},\nonumber\\
f_2(-t\Box) &=& \int_0^1 du u(1-u)\int_0^1 dv e^{u(1-u)v t \Box},\nonumber
\eea
the form factors $\mathscr{F}_i(x)$ can be expressed as (see Ref.~\cite{gusev} for details):
\bea
\mathscr{F}_1(-t\Box) &=& {1\over 12} f_1(-t\Box) - {1\over 2}f_2(-t\Box),\nonumber\\
\mathscr{F}_2(-t\Box) &=& {1\over 12} f_1(-t\Box).\nonumber
\eea
We should notice that the above expression (\ref{zbv}) is valid when the background geometry is asymptotically flat and when the background fields are small in magnitude, but rapidly vary in space. Details can be consulted in Refs.~\cite{BV,gusev}.

The computation of the effective action requires some effort, but it is otherwise straightforward. For convenience, we will write
\bea
\zeta_{\varphi} (s) = \zeta_{\varphi}^{(1)} (s) + \zeta_{\varphi}^{(2)} (s) +\cdots,
\label{zbvgz}
\eea
where $\zeta_{\varphi}^{(k)} (s)$ corresponds to the $k$th term in (\ref{zbv}). The term coming from $k=0$ has been discarded, since it does not depend on the condensate and can be removed by trivial renormalization. The first nontrivial contribution comes from the second term in (\ref{zbv}) and can be written, after simple steps, as
\bea
\zeta_{\varphi}^{(1)} (s) &=& 
{2\beta\over (4\pi)^{D/2}} \sum_{\epsilon=\pm 1} \mathscr{A}_{\epsilon} 
{\Gamma(D/2-1-s)\over\Gamma(s)} \mathscr{O}(s)
\nonumber
\eea
where
\bea
\Re \mathscr{O}(s) &=& 2^{2s -D} \beta^{2s-D+2}\left[
\mbox{Li}_{-+}
+\mbox{Li}_{++}
+ \mbox{Li}_{--}
+\mbox{Li}_{--}
\right],\nonumber
\eea
\bea
\Im \mathscr{O}(s) &=& 2^{2s -D} \beta^{2s-D+2}\left[\mbox{Li}_{-+}-\mbox{Li}_{++}-\mbox{Li}_{--}+\mbox{Li}_{+-}
\right],\nonumber
\eea
with 
\bea
\mbox{Li}_{++}&\equiv&\mbox{Li}_{-2+D-2s}\left(\Phi e^{+\beta\left(\mu +\imath \varphi\right)}\right)~,\nonumber\\
\mbox{Li}_{--}&\equiv&\mbox{Li}_{-2+D-2s}\left(\Phi^*e^{-\beta\left(\mu -\imath \varphi\right)}\right)~,\nonumber
\nonumber
\eea
and defining $\mbox{Li}_{\nu}(z)$ the polylogarithm function. 

Computing the contribution $\zeta_\varphi^{(2)}(s)$ requires some more work. Simple inspection of the form factors shows that
\bea
\zeta_{\varphi}^{(2)} (s) &=& 
{\beta\over (4\pi)^{D/2}} \sum_{\epsilon=\pm 1}
\Big(
\mathscr{A}_\epsilon \mathscr{Y}_1(s) R
+ \mathscr{A}_\epsilon \mathscr{Y}_2(s) \mathscr{A}_\epsilon + \cdots
\Big),\nonumber
\eea
where
\bea
\mathscr{Y}_i (s) &=& {1\over \Gamma(s)} \int_0^{\infty} dt t^{s-D/2+1} \mathscr{F}_i\left(-t\Box\right)
\left(1 +\right.\nonumber\\
&&+\left. 2 \sum_{n=1}^\infty \Upsilon_\varphi^{(n)}\left(\beta,\mu,\varpi\right) e^{-{\beta^2 n^2\over 4t}} \right).
\nonumber
\eea
Direct integration over $t$ can be carried out straightforwardly and allows one to express the functions $\mathscr{Y}_i(s)$ as
\bea
\mathscr{Y}_1 (s) &=& {1\over 12}\mathscr{I} (s) - {1\over 2} \mathscr{J} (s),\nonumber\\
\mathscr{Y}_2 (s) &=& {1\over 12}\mathscr{I} (s),\nonumber
\eea
in terms of 
\begin{widetext}
\bea
\mathscr{I}(s) &=& 
\left(-\Box\right)^{-2-s+D/2} {\Gamma^2\left(
D/2-1-s\right)\Gamma\left(2
-D/2+s\right)\over \Gamma\left(-2
+D-2s\right)\Gamma\left(s\right)}+\nonumber\\
&&+{4\left(-1\right)^{
-D/4+s/2+1}\over \Gamma (s)} 
\left(-\Box\right)^{
D/4-s/2-1} 
\sum_{n=1}^\infty \Upsilon_\varphi^{(n)}\left(\beta,\mu,\varpi\right) \left({n \beta\over 2}\right)^{
-D/2+s+2} \mathscr{K}_n^{(
D)}(s,\Box),
\eea
\bea
\mathscr{J}(s) 
&=&
-{(1+D-2s)\pi^{3/2}\over 2^{D-2s}} 
{\csc\left((D-2s)\pi/2\right)\over \Gamma\left(3/2+D/2-s\right)\Gamma\left(s\right)}\left(- \Box\right)^{D/2-2-s}\nonumber\\
&&-{(-1)^{D/4-s/2-1} \over \Gamma\left(s\right)}\left(-\Box\right)^{D/4-s/2-1}
\sum_{n=1}^\infty \Upsilon_\varphi^{(n)}\left(\beta,\mu,\varpi\right) \left({n \beta\over 2}\right)^{2-D/2+s} \mathscr{R}_n^{(D)}(s,\Box),
\eea
where we have defined
\bea
\mathscr{K}_n^{(
D)}(s,z)&:=&
\int_0^1 du \left(u(u-1)\right)^{
D/4-s/2-1}  K_{
\frac{D}{2}-s-2}\left(n \beta \sqrt{u(u-1)}  \sqrt{x}\right),\nonumber\\
\mathscr{R}_n^{(D)}(s,z)&:=&
\int_0^1 du \int_0^1 dv
\left(u(u-1)\right)^{D/4-s/2} v^{D/4-s/2-1}    
K_{\frac{D}{2}-s-2}\left(n \beta \sqrt{v u(u-1)}  \sqrt{z} \right).\nonumber
\eea
\end{widetext}
The above formulas allow one to obtain the effective action after analytical continuation of $\zeta_\varphi(s)$ and $\zeta_\varphi'(s)$ to $s=0$. Further manipulation of the form factors is possible following some of the procedures outlined in Ref.~\cite{gusev} and we leave them up to the interested reader.

\end{document}